\documentclass[aps,prl,twocolumn,showpacs,superscriptaddress,groupedaddress]{revtex4}

\usepackage{amssymb}
\usepackage{color,graphicx}
\usepackage{amsmath}
\usepackage{amsbsy}
\usepackage{amsthm}
\usepackage{bbm}
\usepackage{bm}
\usepackage{epsfig}
\usepackage{lscape}
\usepackage{float}
\usepackage{subfigure}

\newcommand{\D}{\text{d}}

\begin{document}

\title{Improved Noninterferometric Test of Collapse Models Using Ultracold Cantilevers}

\author{A. Vinante}
\email{andrea.mistervin@gmail.com}
\affiliation{Istituto di Fotonica e Nanotecnologie, CNR - Fondazione Bruno Kessler, I-38123 Povo, Trento, Italy}

\author{R. Mezzena}
\affiliation{Department of Physics, University of Trento, I-38123 Povo, Trento, Italy}
\affiliation{Istituto Nazionale di Fisica Nucleare, TIFPA, Via Sommarive 14, I-38123, Trento, Italy}

\author{P. Falferi}
\affiliation{Istituto di Fotonica e Nanotecnologie, CNR - Fondazione Bruno Kessler, I-38123 Povo, Trento, Italy}
\affiliation{Istituto Nazionale di Fisica Nucleare, TIFPA, Via Sommarive 14, I-38123, Trento, Italy}

\author{M. Carlesso}
\affiliation{Department of Physics, University of Trieste, Strada Costiera 11, 34151 Trieste, Italy}
\affiliation{Istituto Nazionale di Fisica Nucleare, Trieste Section, Via Valerio 2, 34127 Trieste, Italy}

\author{A. Bassi}
\affiliation{Department of Physics, University of Trieste, Strada Costiera 11, 34151 Trieste, Italy}
\affiliation{Istituto Nazionale di Fisica Nucleare, Trieste Section, Via Valerio 2, 34127 Trieste, Italy}

\date{\today}

\begin{abstract}
Spontaneous collapse models predict that a weak force noise acts on any mechanical system, as a consequence of the collapse of the wave function. Significant upper limits on the collapse rate have been recently inferred from precision mechanical experiments, such as ultracold cantilevers and the space mission LISA Pathfinder. Here, we report new results from an experiment based on a high-Q cantilever cooled to millikelvin temperature, which is potentially able to improve the current bounds on the continuous spontaneous localization (CSL) model by one order of magnitude. High accuracy measurements of the cantilever thermal fluctuations reveal a nonthermal force noise of unknown origin. This excess noise is compatible with the CSL heating predicted by Adler. Several physical mechanisms able to explain the observed noise have been ruled out.
\end{abstract}


\maketitle

Spontaneous wave function collapse models~\cite{GRW,CSL,collapse_review1,collapse_review2} are stochastic nonlinear modifications of standard quantum mechanics, which have been introduced as a possible solution of the measurement problem. According to such models, the stochastic collapse of the wave function is a dynamical process which naturally breaks the quantum superposition principle. The process would occur in atomic systems on a very long time scale, practically unobservable, so that standard quantum mechanics would hold strictly. However, the scaling of the collapse rate with the size of the system would lead to a rapid localization of any macroscopic system, and to the emergence of the definiteness of the classical everyday world.

Here we consider the continuous spontaneous localization (CSL) model~\cite{CSL}. CSL is the most known and studied collapse model and has been extensively reviewed in many recent papers \cite{collapse_review1,collapse_review2}. It is characterized by two phenomenological constants, a collapse rate $\lambda$ and a length $r_C$, which characterize, respectively, the intensity and the spatial resolution of the spontaneous collapse.
The conservative value for the collapse rate suggested by Ghirardi, Rimini and Weber \cite{GRW,CSL} is $\lambda \simeq  10^{-16}$ s$^{-1}$ at $r_C=10^{-7}$ m, and is obtained by imposing the collapse to be effective at the macroscopic human scale. A collapse rate $10^{9 \pm 2}$ times larger has been suggested by Adler \cite{adler}, motivated by the requirement of making the wave function collapse effective at the mesoscopic level.

Direct laboratory tests based on quantum superposition experiments set limits on $\lambda$ at the level of $10^{-6}$ s$^{-1}$ \cite{exp_MW,exp_MW2,exp_MW3}. Much stronger bounds can be set by indirect noninterferometric tests based on mechanical systems ~\cite{collett, adler2005, bassi2005,bassi,nimmrichter,diosi,goldwater,vitali}. Relevant mechanical bounds on $\lambda$ at the level of $10^{-8}$ s$^{-1}$ for $r_C = 10^{-7}$m have been recently set by cantilever-based experiments \cite{vinante}, cold atoms \cite{kasevich,atomi} and the space-based experiment LISA Pathfinder \cite{vitale, LISA, helou}. Stronger bounds, though less robust to variations on the model \cite{adlerX, donadiX}, are set by spontaneous emission of x rays \cite{curceanu}.

Here, we report on a improved version of the cantilever experiment, with a relative strength of the thermal noise force reduced by 1 order of magnitude. Unlike the previous cantilever experiment \cite{vinante}, we find evidence of a nonthermal excess noise of unknown origin. If interpreted as CSL-induced noise, this would be compatible with previous experimental bounds and in agreement with the collapse rate predicted by Adler \cite{adler}. Alternatively, if the noise finds an explanation within standard physics, its identification and elimination will allow us to extend the experimental bound on $\lambda$ by 1 order of magnitude, almost ruling out Adler's proposal.

\begin{figure}[!ht]
\includegraphics{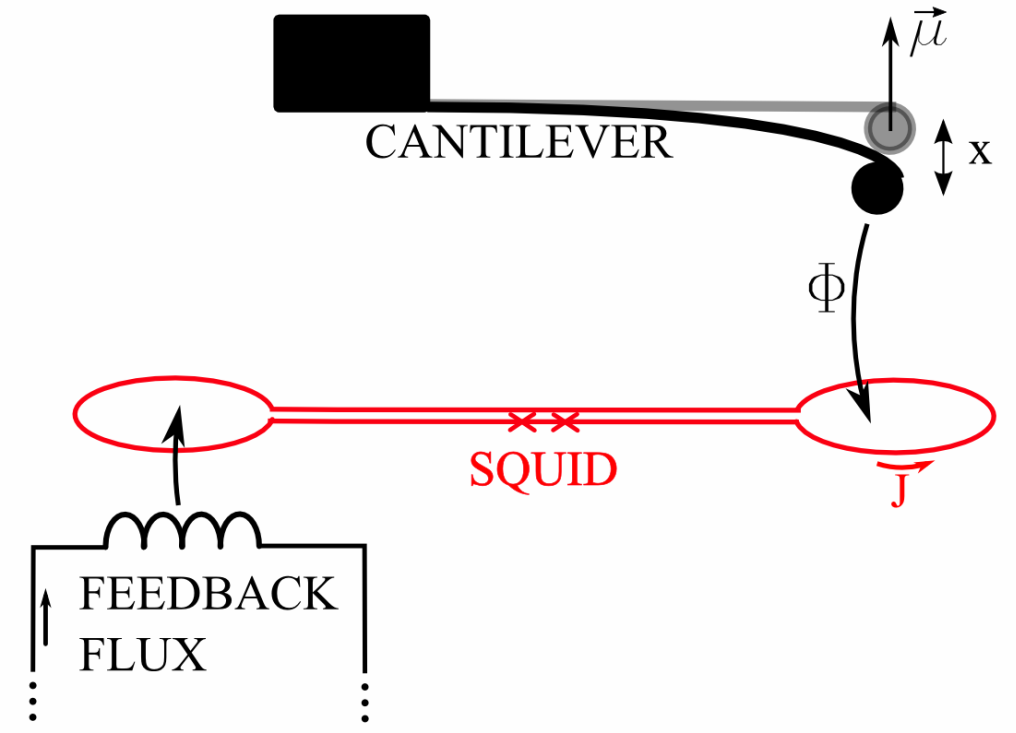}
\caption{Simplified measurement scheme. The fundamental bending mode of a cantilever loaded with a ferromagnetic microsphere with magnetic moment $\mu$ is continuously monitored by a SQUID susceptometer. The SQUID measures the magnetic flux $\Phi=\Phi_x x$ coupled by the displacement $x$ of the magnetic particle and is operated in flux-locked-loop (the feedback electronics is not shown for simplicity). The flux-dependent circulating current $J$, combined with finite feedback gain, causes a dynamical magnetic spring effect that modifies the apparent quality factor of the cantilever.}  \label{fig1}
\end{figure}

The scheme of our experiment is similar to the one described in Ref. \cite{vinante} and is shown in Fig. \ref{fig1}. A cantilever with a spherical ferromagnetic load is continuously monitored by a superconducting quantum interference device (SQUID). The displacement $x$ is converted into magnetic flux by a linear coupling $\Phi_x=d\Phi/dx$ which depends on the magnet position and orientation. A novel feature is that the mechanical quality factor $Q$ is much higher than in previous experiments, and heavily temperature dependent. Moreover, we observe a dynamical SQUID-induced magnetic spring effect, analog to optical spring effects in optomechanics, which modifies the quality factor from its intrinsic value $Q$ to an apparent value $Q_a$ \cite{clarke1}. To account for this new feature, we seek a strategy to directly measure the effective force noise acting on the resonator, rather than the mean energy which may be affected by the dynamically modified quality factor $Q_a$.
To this end, let us consider the Lorentzian spectral density associated with the cantilever displacement fluctuations
\begin{equation}
S_x  = \left( \frac{S_{F0}}{k^2} + \frac{{4k_B T}}{{k\omega _0 Q}} \right) \frac{{f_0 ^4 }}{{\left( {f_0 ^2  - f^2 } \right)^2  + \left( {\frac{{ff_0 }}{{Q_a }}} \right)^2 }}   \label{Sx}
\end{equation}
Here, $f_0=\omega_0 / 2 \pi$ is the resonant frequency, $k$ is the spring constant, $k_B$ is the Boltzmann constant, $T$ is the temperature and $S_{F0}$ is the spectral density of any nonthermal force noise. Equation~(\ref{Sx}) says that magnetic spring effects, similar to optomechanical ones, only affect the dynamics and thus the denominator of the resonant term, characterized by an apparent quality $Q_a$ \cite{aspelmeyerreview, pootreview, falferiIBM}. Instead, the amplitude of the Lorentzian curve is proportional to the total force noise. The thermal contribution scales with $T/Q$, where $Q$ is the intrinsic quality factor. Our strategy is to characterize the thermal noise by accurate measurements as function of $T/Q$. Any excess nonthermal noise, included that caused by CSL, will cause a constant force spectral density $S_{F0}$, independent of $T/Q$. The maximum nonthermal force noise compatible with the experiment can be used to test CSL predictions. This requires modeling the CSL force acting on the continuous mechanical resonator, exactly as done in Ref.~\cite{vinante} (see Supplemental Material for details \cite{supplemental}).

The mechanical resonator in our setup is a commercial tipless AFM silicon cantilever, with size $450 \times 57 \times 2.5$ $\mu$m$^3$ and stiffness $ k= \left( 0.40 \pm 0.02 \right)$ N/m. A hard ferromagnetic microsphere (radius $R=15.5$ $\mu$m, density $\rho=7.43$ kg/m$^3$) is glued to the cantilever free end and magnetized. The microsphere has a twofold function. It increases the cross section to the CSL field, which scales as $\rho^2$ \cite{vinante}, while at the same time enabling a straightforward detection by means of a nearby SQUID susceptometer. The SQUID is gradiometric and comprises two distant loops with radius $R_S=10$ $\mu$m \cite{vinante2}. The particle is aligned above the first loop at a height $h \simeq 40$ $\mu$m, with the motion of the first flexural mode orthogonal to the SQUID plane. The SQUID is operated in two-stage flux-locked-loop configuration with the feedback applied to the second loop (see Fig. 1). This geometry strongly suppresses direct coupling between the feedback signal and the cantilever.

The cantilever-SQUID system is enclosed in a copper box, suspended above the mixing chamber plate of a pulse-tube dilution refrigerator (Janis Jdry-100-Astra) by means of a two-stage suspension system. The measured mechanical attenuation is higher than $80$ dB at the cantilever frequency. The temperature of the mixing chamber is measured by a RuO$_2$ thermometer, while the temperature of the SQUID box is measured by a SQUID noise thermometer. Both devices have been calibrated against a superconducting reference point device with accuracy better than $0.5 \%$.

The resonant frequency of the fundamental mode of the cantilever is $f_0=8174.01$ Hz. The actual mechanical quality factor $Q_a$ is determined by ringdown measurements. The dynamical magnetic spring effect depends on the SQUID working point and scales as $1/|G|$ where $G$ ($|G|\gg 1$) is the open loop gain of the feedback electronics. Specifically, we expect a linear dependence of the actual quality factor $Q_a$ as $1/Q_a=1/Q+c/|G|$, where $c$ is a constant depending on the SQUID working point. 
This behavior is experimentally observed by varying the gain $|G|$, and allows us to infer the intrinsic quality factor $Q$. The whole measurement procedure is repeated at each temperature $T$ \cite{supplemental}.

The measured intrinsic $Q$ is of the order of $6\times 10^5$ at temperature of the order of $1$ K, and is observed to increase roughly as $1/T$ upon reducing the temperature below $T=500$ mK, approaching $Q \simeq 10^7$ at $T\simeq 20$ mK. This behaviour is consistent with measurements performed with the SQUID weakly coupled, and is reminiscent of two-level systems dissipation \cite{parpia}. Standard glassy two-level systems in a $2$ nm amorphous oxide layer on the cantilever surface are able to explain the observed effect quantitatively.

The noise is measured by acquiring and averaging spectra of the SQUID signal, calibrated as magnetic flux, with typical integration time of $800-1200$ s. Before measuring the noise and the quality factor, the system is allowed to thermalize for at least 3 hours. During the noise measurement the pulse tube is switched off, and the mixing chamber temperature actively stabilized by a PID controller.
\begin{figure}[!ht]
\includegraphics{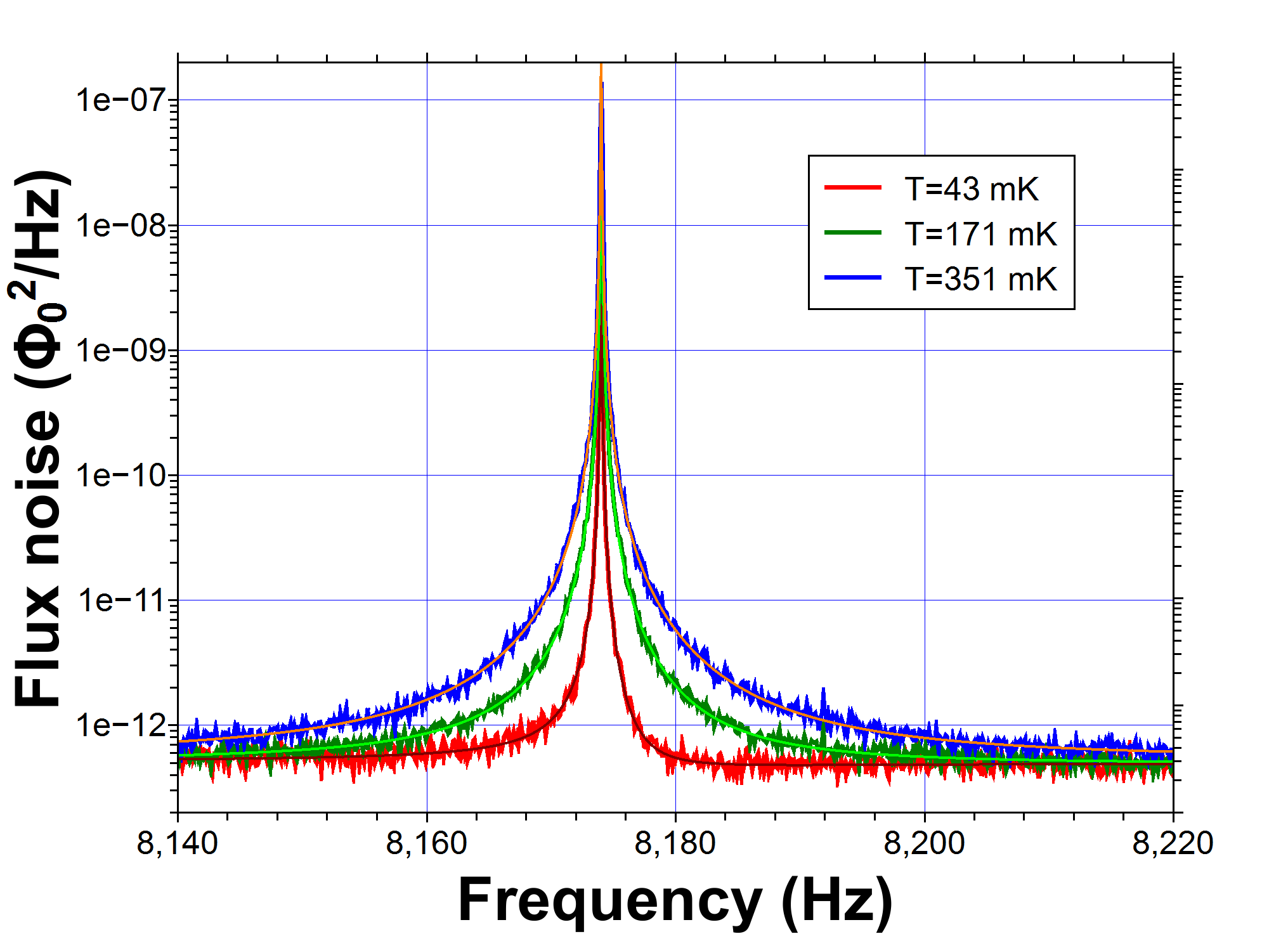}
\caption{Examples of averaged spectra at three representative temperatures, with the respective best fit with Eq.~(\ref{fit}). }  \label{spectra}
\end{figure}
Examples of averaged flux noise spectra at three representative temperatures are shown in Fig. \ref{spectra}. 
The spectra are fitted with the curve
\begin{equation}
S_\Phi   = A + \frac{{Bf_0 ^4  + C\left( {f^2  - f_1 ^2 } \right)^2 }}{{\left( {f^2  - f_0 ^2 } \right)^2  + \left( {\frac{{ff_0 }}{{Q_a }}} \right)^2 }}
  \label{fit}
\end{equation}
The term proportional to $B$ is the relevant one, as it corresponds to the fluctuations of the cantilever induced by thermal or extra force noise, given by Eq.~(\ref{Sx}), converted into magnetic flux. The term proportional to $A$ is the purely additive wideband noise of the SQUID. The last term, proportional to $C$ arises because of the flux noise applied to the SQUID by the feedback electronics in order to compensate for the SQUID additive wideband noise. This flux noise induces a current $J$ circulating in the SQUID through the finite responsivity $J_\Phi=dJ/d\Phi$ \cite{clarke1}, which eventually leads to an effective backaction on the cantilever. The transfer function of this mechanism has been directly measured by injecting a calibration signal and features an antiresonance at $f=f_1$, with $f_1-f_0=1.1$ Hz. The overall effect is a small asymmetric distortion of the Lorentzian peak. 

All spectra have been checked by $\chi^2$ tests \cite{supplemental} to be consistent with Eq.~(\ref{fit}). All estimations of the SQUID parameters $A$ and $C$ are consistent with each other, with mean values $A=\left( 1.23 \pm 0.05 \right) \times 10^{-13}$ $\Phi_0^2$/Hz and $C=\left( 3.78 \pm 0.05 \right) \times 10^{-13}$ $\Phi_0^2$/Hz. In particular, $A$ and $C$ do not depend significantly on temperature. This is expected, as for this type of SQUID \cite{falferiIBM} the noise is saturated by hot electron effect \cite{wellstood} for $T<400$ mK, and our measurements satisfy this condition.

\begin{figure}[!ht]
\includegraphics{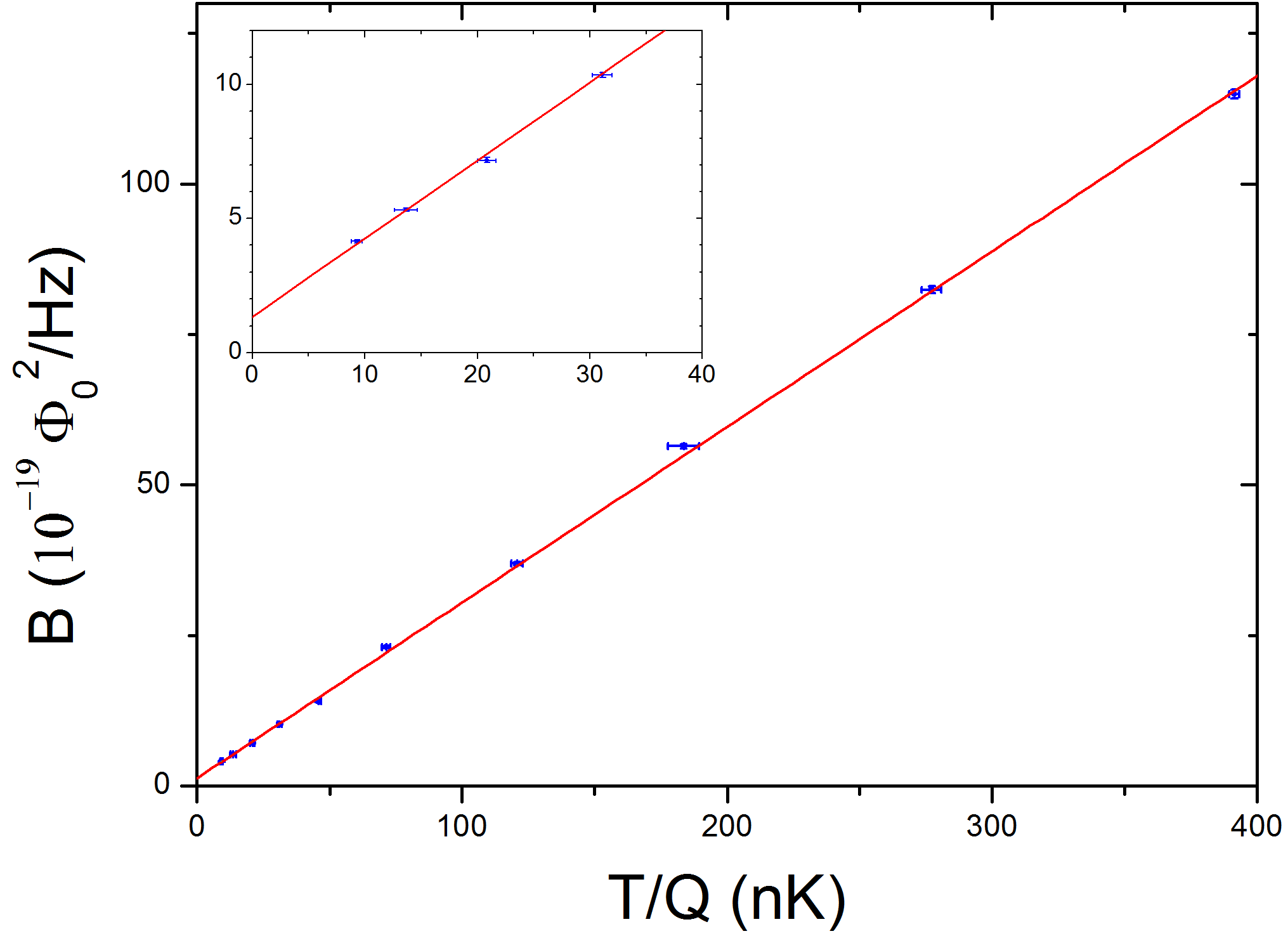}
\caption{Symmetric amplitude of the Lorentzian noise $B$, as measured by the SQUID, as function of the ratio  $T/Q$, together with the best linear fit. In the inset, the data at the lowest $T/Q$ are zoomed in order to highlight the nonzero intercept of the fit. }  \label{TvsT}
\end{figure}

Figure~\ref{TvsT} shows the measured symmetric amplitude $B$ of the Lorentzian noise as function of $T/Q$, varied by changing the bath temperature. The uncertainty on the estimation of $B$ is remarkably low, of the order of $1 \%$. The $x-$error bar, dominated by the uncertainty on $Q$, is thus significant \cite{supplemental}.
The data agree with a linear behavior over the whole $T/Q$ range, in agreement with Eq.~(\ref{Sx}). A weighted orthogonal linear fit with the expression $B_0+B_1 T/Q$ yields the intercept $B_0=\left( 1.27 \pm 0.11 \right)\times 10^{-19}$ $\Phi_0^2/$Hz and the slope $B_1= \left( 0.291 \pm 0.002 \right) \times 10^{-19}$ $\Phi_0^2/ \left( \mathrm{nK} \cdot \mathrm{Hz} \right)$. In addition, we exploit the linear dependence on $T/Q$ to infer the coupling between cantilever and SQUID. Given Eqs.~(\ref{Sx}) and (\ref{fit}) we can express the thermal slope $B_1$ as
\begin{equation}
B_1  = \frac{{4k_B }}{{\omega _0 }}\frac{{\Phi _x ^2 }}{k} ,   \label{B_1}
\end{equation}
which allows the coupling factor $\Phi_x^2/k$ to be evaluated from the measured $B_1$.

The finite intercept, clearly visible in the inset of Fig. \ref{TvsT} implies that the data are not compatible with a pure thermal noise behavior, and a nonthermal excess noise is present. According to Eq.~(\ref{Sx}) we can convert $B_0$ into a residual force noise
\begin{equation}
S_{F0}  = \frac{{4k_B k}}{{\omega _0 }}\frac{{B_0 }}{{B_1 }} .    \label{SF0}
\end{equation}
The measured coupling factor and residual force noise are reported in Table \ref{tabella}, together with the same quantities inferred from the additional measurements discussed in the following. The systematic error on $S_{F0}$ arises from the uncertainty on $k$.

We checked for possible physical sources of the excess force noise. First, we expect a backaction force spectral density $S_{F,BA}=S_J F_J^2$ from the noise in the current circulating in the SQUID loop. Here, $S_J$ is the current spectral density and $F_J=dF/dJ$ is the backward current-to-force factor. Because of reciprocity, $F_J$ must be equal to the forward displacement-to-flux factor $\Phi_x$ \cite{usenko} so that $S_{F,BA}=S_J \Phi _x ^2$. In other words, the backaction noise leads to a finite intercept, and the corresponding force noise scales with the coupling factor. 
\begin{table}
\begin{center}
\begin{tabular}{|c || c | c |}
\hline
Pulse tube &  $\Phi_x^2/k$ (fH)  & $S_{F0}$ (aN$^2$/Hz) \\
\hline 
Off  & $116 \pm 1$ &  $1.87 \pm 0.16 \pm 0.1 $(sys)  \\
\hline
Off & $347 \pm 3$ & $2.12 \pm 0.20 \pm 0.1$(sys) \\
\hline
On & $114 \pm 2$ & $2.58 \pm 0.20 \pm 0.1 $(sys)   \\
\hline   
\end{tabular}
\end{center}
\caption{Operating conditions, measured coupling and residual force noise for the different measurement data sets. } \label{tabella}
\end{table} 

We took advantage of this property and performed additional measurements in a subsequent cooldown at a different cantilever position, with effective coupling increased by a factor of $\sim 3$. We observe again a linear behavior in very good agreement with the experimental data, with a finite intercept \cite{supplemental}. The corresponding residual force noise is reported in the second row of Table \ref{tabella} and is consistent within the error bar with the one at low coupling. This clearly indicates that most of the observed excess noise cannot be attributed to SQUID backaction.

We can compare this result with the prediction of the Clarke-Tesche model \cite{clarke2} for the current noise, $S_J=\gamma k_B T_{SQ}/R_{SQ}$ with $\gamma\simeq  11$ for an optimized SQUID. Here $R_{SQ}=8$ $\Omega$ and $T_{SQ}\simeq 400$ mK are the measured shunt resistor and the typical SQUID electron temperature. From this expression we estimate a small increase of the backaction force noise $\Delta S_{F,BA} \simeq  0.6$ aN$^2$/Hz between the two measurements, which is compatible within $2\sigma$ with the experimental increase $\Delta S_{F0}=\left( 0.24 \pm 0.26 \right)$ aN$^2$/Hz.

In order to investigate the role of vibrational noise from the refrigerator or from the outside world, we repeated the measurements at low coupling by keeping the pulse tube on \cite{supplemental}. The input mechanical noise provided by the pulse tube in our cryostat is known to be more than 2 orders of magnitude larger than the background noise when the pulse tube is off. However, while the measured spectra with the pulse tube on are significantly dirtier, we can still perform a Lorentzian fit and the residual force noise, reported in Table \ref{tabella}, is only slightly increased with respect to the measurements with the pulse tube off. This confirms that the mechanical suspensions are working well within design specifications, and suggests that vibrational noise is not the source of the observed excess noise with the pulse tube off. We have also ruled out vibrational noise from the $^3$He flow, by switching the circulation pump on and off without noticeable effects \cite{supplemental}.

Magnetic effects, such as fluctuations of the environmental magnetic field or fluctuations of the microsphere magnetization, can be also considered as possible excess noise sources. We can substantially rule out these mechanisms, based on theoretical order-of-magnitude estimations, and a further test which has shown the quality factor to be independent of the external static field \cite{supplemental}.

Another option is that we are actually observing thermomechanical noise, but the effective temperature of the noise source (or part of it) is higher than the one of the thermal bath because of thermal gradients along the cantilever. In this case one would expect to observe saturation effects, as observed in \cite{vinante} rather than a linear behavior with a fixed intercept. Furthermore, we have performed simple thermal modeling of the cantilever. The power dissipated in the magnet by eddy currents induced by SQUID Josephson radiation is estimated to be of the order of 1 fW, and would cause a temperature gradient between the magnet and the cantilever base smaller than $1$ mK in the temperature range explored by this experiment.

The observed finite intercept could also be a subtle artifact due to an unknown systematic error in the determination of $1/Q$. We find that this is in principle possible, but the systematic error on $1/Q$ would have to be $10$ times larger than the statistical error bar to be consistent with zero excess noise. Moreover, the data would no longer follow a linear behavior \cite{supplemental}.

\begin{figure}[!t]
\includegraphics{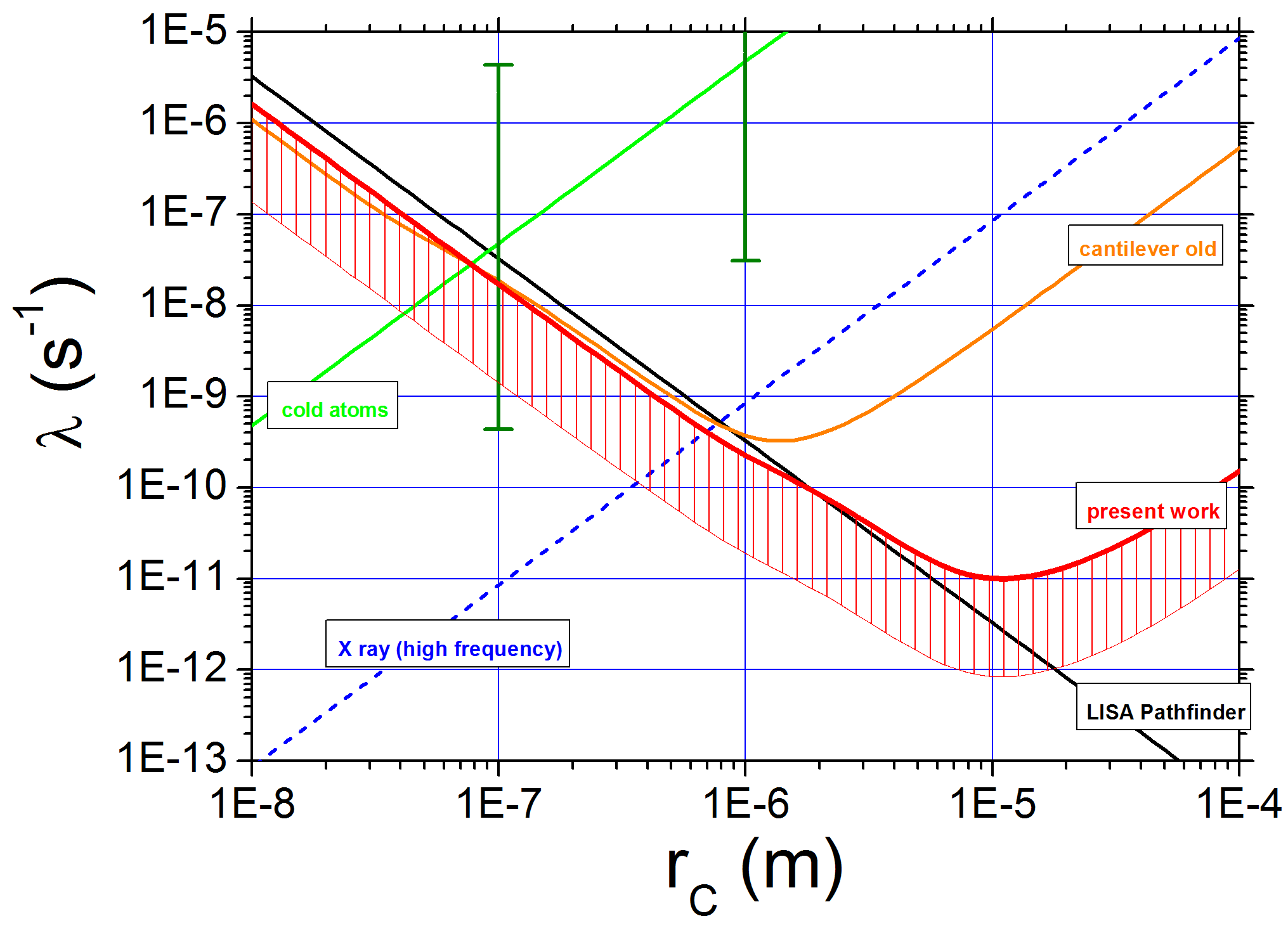}
\caption{Exclusion plot in the $\lambda - r_C$ plane based on our experimental data, compared with the best experimental upper bounds reported so far and with theoretical predictions. Continuous thick (red) curve: CSL collapse rate $\lambda$, as function of the characteristic length $r_C$, assuming that the observed noise is entirely due to CSL. The shaded region would be excluded by our experiment if the physical origin of the excess noise were identified. The other thin lines represent upper limits from labeled experiments: previous cantilever experiment (orange) \cite{vinante}, LISA Pathfinder (black) \cite{LISA}, cold atoms (green) \cite{atomi} and x-ray spontaneous emission (dashed blue) \cite{curceanu}. The (dark green) bars represent the CSL collapse rate suggested by Adler \cite{adler}.}   \label{upperlimit}
\end{figure}

Finally, let us compare our results with the predictions of the CSL model. By using the same method discussed in Ref. \cite{vinante}, we can convert the observed excess noise into the red curve in the parameter space $\lambda-r_C$ of the CSL model, shown in Fig. \ref{upperlimit}. This curve can be considered as a conservative improved upper bound from mechanical experiments for $r_C \in [ 10^{-7},2\times10^{-6} ]$ m. If the excess noise were indeed due to CSL, the true CSL parameters would actually lie somewhere on the curve. For the standard choice $r_C=10^{-7}$ m this would imply $\lambda=10^{-7.7}$ s$^{-1}$, in agreement with Adler's predictions \cite{adler}. Alternatively, if the observed noise can be eventually reduced to standard physical effects, its identification and elimination will lead to an improved upper bound on CSL, determined by the experimental error bar. The parameter region which can be potentially excluded is shaded in Fig. \ref{upperlimit}. A full exclusion would almost completely rule out Adler's predictions \cite{adler}.

In conclusion, we have performed an improved cantilever-based test of the CSL model. The new experiment features excess noise, which is in principle compatible with the predictions by Adler \cite{adler}. Several physical mechanisms able to explain the observed excess noise have been ruled out. Further investigations are needed in order to probe other possible explanations. Besides further analysis of the present experiment, it will be important to repeat the experiment with a modified setup, for instance with a different cantilever or microsphere, and other groups should possibly repeat similar measurements with a different setup.
Above all, this experiment neatly illustrates the fundamental challenge of collapse model testing. Negative results are robust, but positive claims require extremely careful and systematic work in order to exclude any conceivable alternative physical explanation.

\textit{Acknowledgments --} A.V. acknowledge partial support from the International Centre for Theoretical Studies (ICTS) for the participation to the program - Fundamental Problems in Quantum Physics (Code: ICTS/Prog-fpqp/2016/11), and thanks many participants for useful discussions, in particular Hendrik Ulbricht, Tejinder Singh, Tjerk Oosterkamp, Nikolai Kiesel, Saikat Ghosh and Daniel Sudarsky. A.B. and M.C. acknowledge financial support from the University of Trieste (FRA 2016) and INFN.

\pagebreak 

\section{Supplemental Material}

\setcounter{equation}{0}
\setcounter{figure}{0}
\renewcommand{\theequation}{S\arabic{equation}}
\renewcommand{\thefigure}{S\arabic{figure}}

\section{CSL force noise}

The expression for the CSL-induced force noise spectral density can be obtained directly from the correlations of the CSL force on the system [17,21]. In terms of the CSL parameters and the mass density distribution $\mu(\bm r)$ of the system, the one-sided spectral density $S_{F}$ reads:
\begin{equation}
S_{F}=\frac{2 \hbar^2\lambda r_C^3}{\pi^{3/2}m_0^2}\int\D\bm k\,k_x^2e^{-k^2 r_C^2}|\tilde\mu(\bm k)|^2,
\end{equation}
where $\tilde \mu(\bm k)$ is the Fourier transform of $\mu(\bm r)$ and $x$ is the direction of the monitored oscillations of the system.

In Ref.~[17], a normalized diffusion constant $\eta$ was derived, which coincides with the force noise apart from a constant factor, according to the relation $S_F = 2 \hbar^2 \eta$. Here, the factor $2$ originates from the fact that in the experiment we use the one-sided definition of spectral density. For the cantilever-microsphere system the integration can be carried out exactly as done in Ref.~[17], as the geometry is the same. For a given measured residual force noise $S_{F0}$ the exclusion plot in the CSL parameter space is obtained by comparing $S_{F0}$ with Eq.~(S1).

\section{Additional information on fitting the noise spectra}

The power spectra of the SQUID output signal are experimentally obtained by averaging a number, typically $n_{\mathrm{av}}=120$, of power FFT periodograms of the SQUID signal. The sampling frequency was set to $f_s=100$ kHz and the length of each dataset was $2^{20}$ samples, corresponding to a frame period of $t_f=10.49$ s and a frequency resolution $df=95.36$ mHz.

Each averaged spectrum is fitted with a weighted nonlinear Levenberg-Marquardt procedure on the fixed interval $8100-8240$ Hz, using Eq.~(2) as template and setting the relative error bar of each bin as $1/\sqrt{n_\mathrm{av}}$. As it is usually difficult to fit very narrow resonance peaks, we fix the parameters $f_0, f_1$ and $Q$ to the values independently determined by ringdown and calibration measurements, leaving only $A$,$B$ and $C$ as free parameters. $A$ and $C$ are mainly determined by the SQUID noise, while $B$ is mainly determined by the tails of the resonant peak.
\begin{figure}
\includegraphics{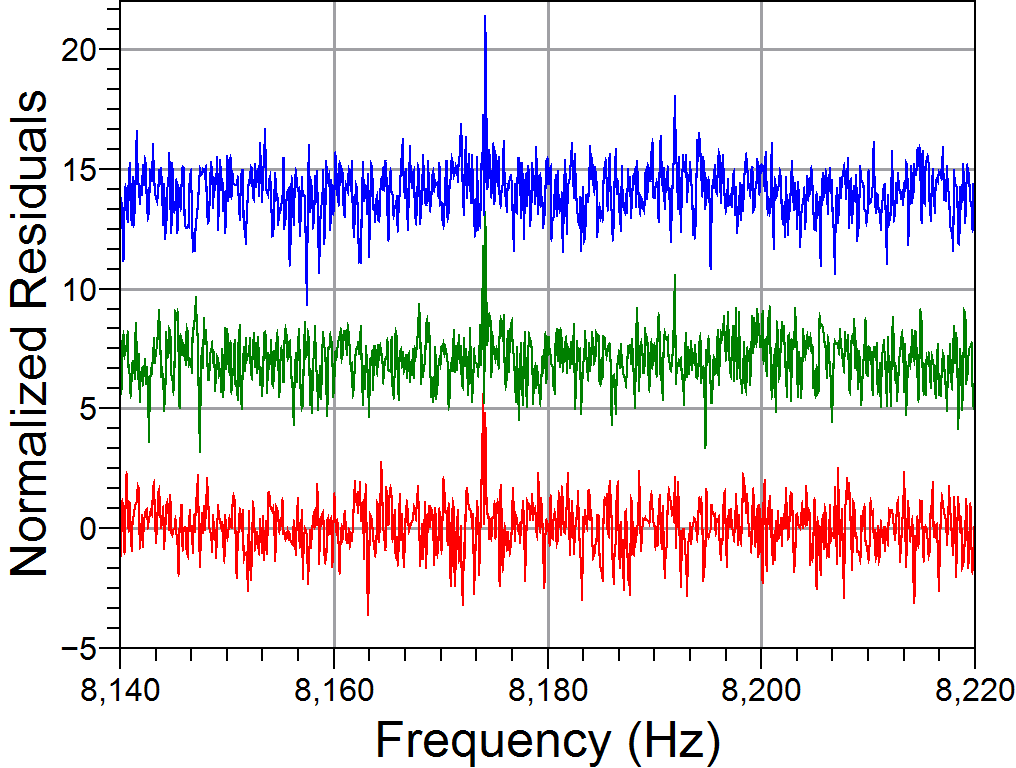}
\caption{Normalized fit residuals for the three representative spectra shown in Fig.~2. From bottom to top, $T=43, 171, 351$ mK respectively. The residuals at $171$ and $351$ mK are shifted from $0$ for a better visualization. \label{figS1}}  
\end{figure}

The fits are typically good, according to a standard $\chi^2$ test. In Fig.~\ref{figS1} we plot the residuals of the fits, normalized by the error bar, for the three representative spectra shown in Fig.~2. In general, no systematic discrepancy between the fitted function and the data is observed, except for a small reproducible imperfect fitting exactly around the resonance frequency of $8174.0$ Hz. This feature can be explained as a data processing artifact due to spectral leakage. In fact, the width of the resonance peak $\Delta f=f_0/Q_a\simeq 50$ mHz is comparable to the spectrum resolution, so that there is a slight broadening, typically non Lorentzian, of the $3-5$ points at the very top of the peak. This localized imperfection leads to a slight increase of the value of $\chi^2$, however it does not significantly affect the fit results. In fact, with the parametrization of Eq.~(2), the $B$ parameter is determined by the whole tails of the resonant peak, while the top part is determined by the fixed $Q_a$ factor. In other words, by fitting with Eq.~(2) we use the information on the cantilever noise available over the whole noise bandwidth (roughly $20$ Hz). This would not be the case for a measurement strategy aiming at measuring the total integrated cantilever noise.

\begin{figure}
\includegraphics{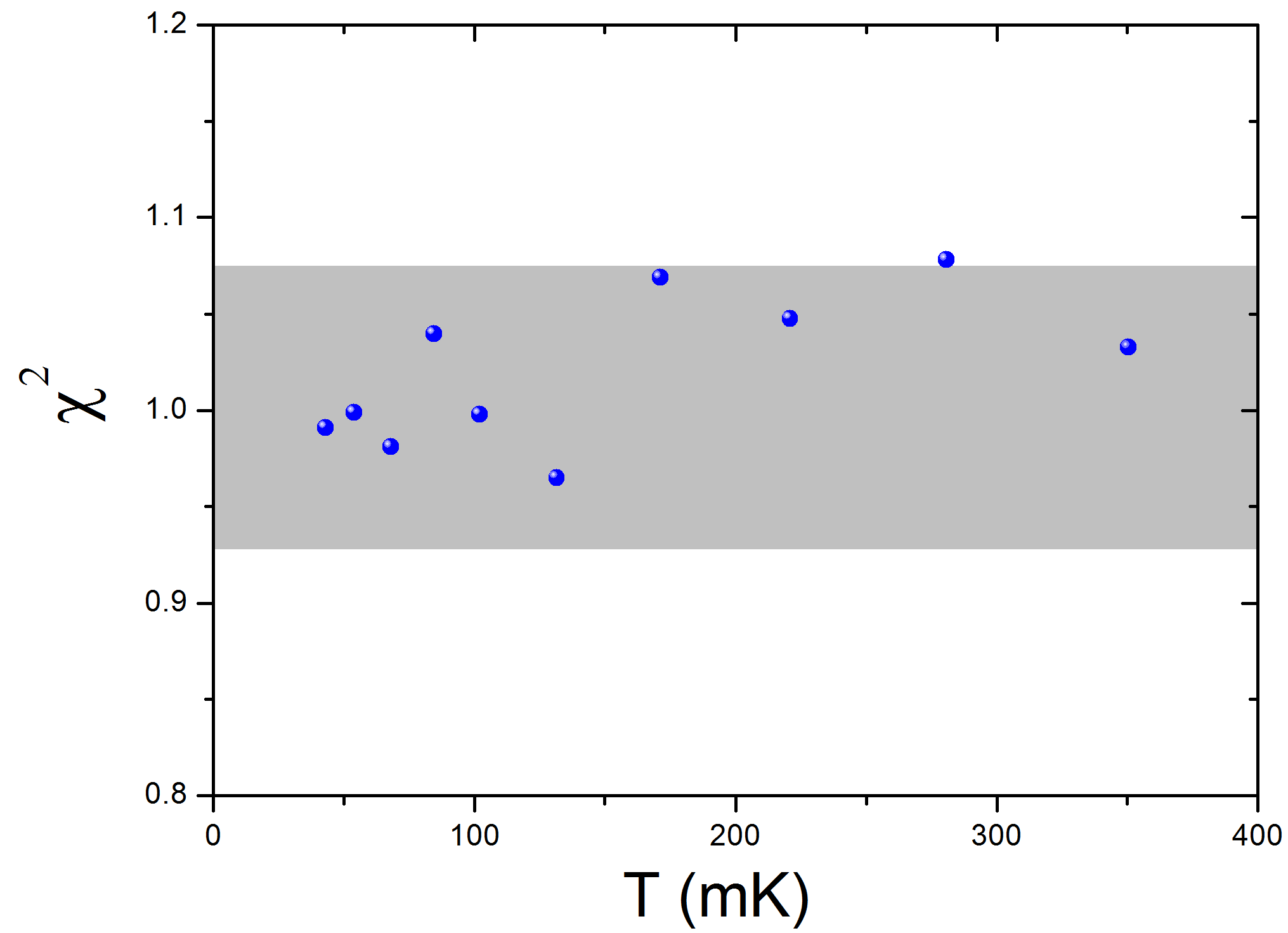}
\caption{Reduced $\chi^2$ of the fits, compared with the theoretical two-sided $2 \sigma$ interval (light gray region) of the reduced $\chi^2$ distribution with the same number of degrees of freedom. \label{figS2}}  
\end{figure}

In Fig.~\ref{figS2} we plot the reduced $\chi^2$ (with 1460 degrees of freedom) of all fits, obtained after exclusion of the $5$ points around the resonance. The gray region represents the theoretical two-sided $2\sigma$ interval of the reduced $\chi^2$ distribution. The experimental $\chi^2$ is essentially in agreement with the theoretical distribution, with only one point slightly beyond the $2\sigma$ limit.

\section{Estimation of the intrinsic quality factor}

The knowledge of the intrinsic quality factor $Q$ is essential in order to evaluate the factor $T/Q$ and therefore to predict the thermal force noise at a given temperature. Unfortunately, we have direct experimental access only to the apparent quality factor $Q_a$, which is affected by the SQUID-induced magnetic spring. In general, we have $1/Q_a=1/Q+1/Q_{SQ}$, where $1/Q_{SQ}$ represents the SQUID effect. It is theoretically expected that, in the limit of large feedback gain:
\begin{equation}
 1/Q_{SQ} =  c/|G|  \label{S2}
\end{equation}
where $|G|$ is the open loop gain of the SQUID feedback electronics, and $c$ is a coupling constant depending on the SQUID working point. This relation is easily derived by noting that the magnetic spring $k_{SQ}$ induced by the SQUID arises from the effective flux $\Phi=\Phi_x x$ applied to the SQUID by the cantilever motion, which in turns generates a current $J$ circulating around the SQUID loop through the responsivity $J_{\Phi}=\frac{dJ}{d\Phi}$ and a back-action force through the coupling $F_J=dF/dJ$. In absence of flux feedback, $k_{SQ}$ can be written as :
\begin{equation}
 k_{SQ}=\frac{dF}{dx}=F_J J_{\Phi} \Phi_x = J_{\Phi} \Phi_x^2   \label{S3}
\end{equation}
where $F_J=\Phi_x$ because of reciprocity, and $J_{\Phi}$ is the only quantity depending on the SQUID working point.
When feedback is applied, the effective flux applied to the SQUID, and therefore the magnetic spring, is reduced by a factor $1/\left[ 1+G \left( \omega \right) \right] $ where $G \left( \omega \right)$ is the open loop gain. In general $G \left( \omega \right)$ is complex, so that the spring features both a real and an imaginary part, leading respectively to a frequency shift $\Delta f_{SQ}$ and a dissipation $1/Q_{SQ}$. Both components are proportional to $1/|G|$ for large $|G|$ and fixed argument. Thus, by varying the magnitude $|G|$ it is possible to distinguish $1/Q_{SQ}$, according to Eq.~(\ref{S2}). In particular, for $1/|G| \rightarrow 0$ the magnetic spring vanishes, allowing the intrinsic quality factor $Q$ to be estimated.

\begin{figure}
\includegraphics{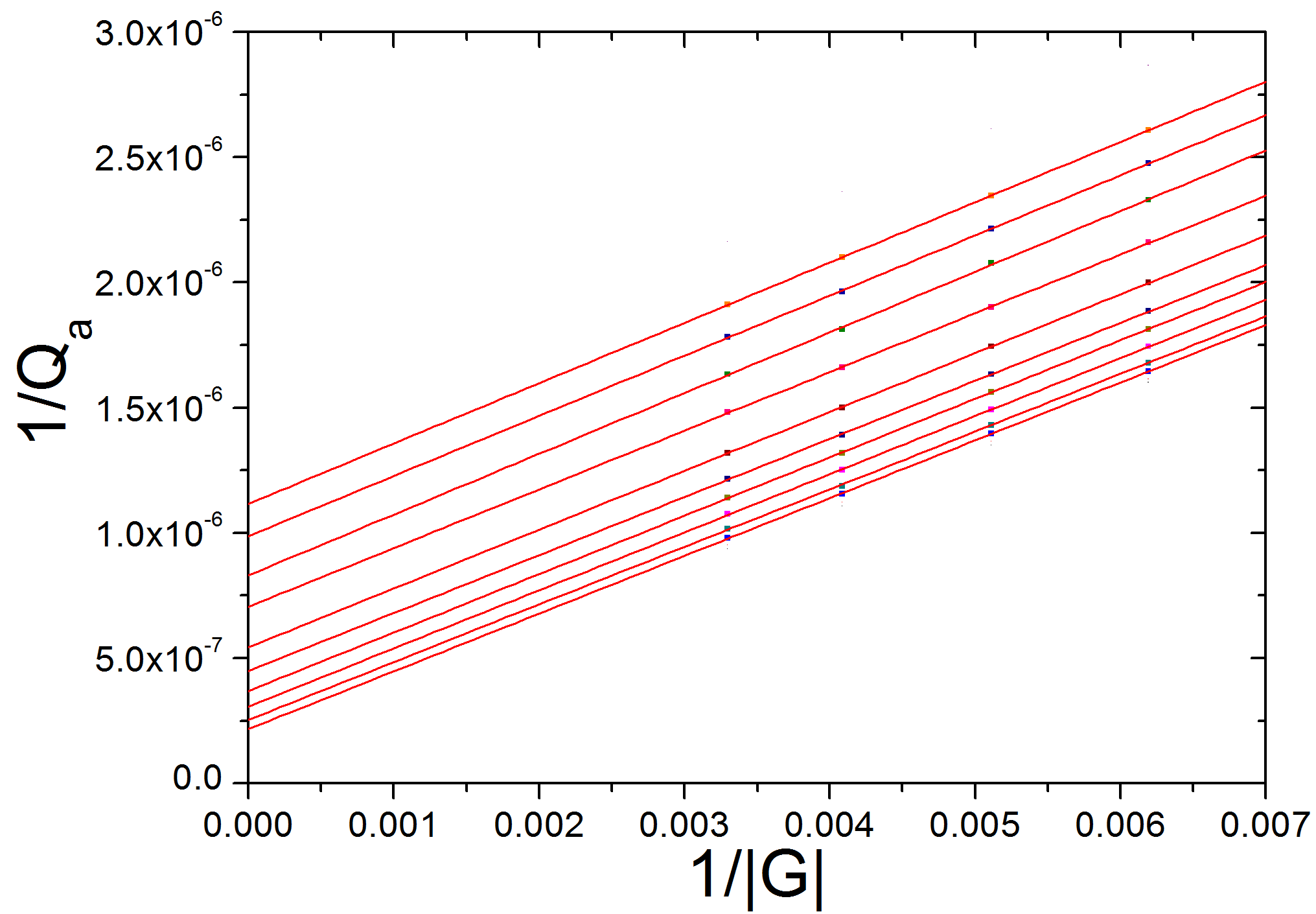}
\caption{Inverse of the apparent quality factor $Q_a$, estimated by ringdown measurements, as function of the inverse of the open loop gain $|G|$ of the SQUID feedback electronics. Each dataset refers to a different temperature: from top to bottom 43,54,68,84,102,132,171,221,281,351 mK. These temperatures correspond to the noise measurements shown in Fig.~3. The intercept of each linear fit provides an estimate of the intrinsic quality factor at the corresponding temperature. \label{figS3}}  
\end{figure}

The apparent quality factor $Q_a$ is measured by using a standard ringdown method. Fig.~\ref{figS3} shows the measurements of $Q_a$ as function of $1/|G|$ at different temperatures, corresponding to the data points of Fig.~3. We could not measure $Q_a$ at larger gain (lower $1/|G|$) because of the onset of the principal feedback instability. All datasets are in very good agreement with the expected linear behaviour. Moreover, all slopes obtained from a linear fit are consistent within the error bar, as expected given that the SQUID working point was the same across all measurements. 

\section{Uncertainty budget on $T$ and $Q$ and possible systematic errors}
The quality factor $1/Q$ at a given temperature is estimated as the intercept of a linear fit to the datasets in Fig.~\ref{figS3}. The error bar is obtained from the standard error on the fitting parameter. Because of the low number of points, the error bar is enlarged by a factor $1.32$, corresponding to the $1\sigma$ (i.e. $68\%$ probability) confidence interval of a Student's $t$-distribution with $2$ degrees of freedom.

The measurement of $T$ is based on a SQUID-based noise thermometer, which has been further calibrated against a superconducting reference point thermometer with accuracy better than $0.5 \%$ (HDL1000 Measurement System, http://hdleiden.home.xs4all.nl/srd1000). The noise thermometer is semiprimary (as it needs only one calibration point) and is simultaneously consistent with all reference points in the range $21$ mK - $1.1$ K, so its effective accuracy could be even better, but we take as conservative accuracy the value $0.5 \%$ set by the reference point device.

The x-error bars on $T/Q$ in Fig.~3 (and in Fig. S5, S7) are obtained by combining in quadrature the relative error on $1/Q$ with the estimated accuracy $\delta T/T=0.005$ on the measurement of $T$. The uncertainty on $T$ is practically negligible with respect to the uncertainty on $Q$.
As the uncertainties on the $x$-axis and the $y$-axis in Fig.~3 are both significant, the linear fit of the noise data is performed as a weighted orthogonal fit which takes simultaneously into account both uncertainties. The goodness of the fit is checked by means of a standard $\chi^2$-test, which gives a regular value $\chi^2=9.27$ with $8$ degrees of freedom. This is a good indication that the error bars are correctly estimated.

Concerning the estimation of $1/Q$ described above, any possible systematic error on the measurement procedure based on varying the open loop gain would be the same for any dataset, as all ringdown measurements of $Q_a$ are performed in the same way at the same settings of the SQUID electronics. One may ask whether a common unknown constant bias on $1/Q$ would be potentially able to explain the nonzero intercept in the noise data of Fig.~3. To check this possibility, we have manually added a constant additional offset $1/Q_0$ to the data, and repeated the whole data analysis. It turns out that it is indeed possible to reduce the intercept to $0$ for $1/Q_0 \simeq  1.1 \times 10^{-7}$, which is roughly $10$ times larger than the average error bar on $1/Q$. However, for such a choice, the data deviate significantly from the linear behaviour, as we obtain $\chi^2=26$ with a relative probability of $0.001$. Furthermore, we find that by varying $1/Q_0$ the $\chi^2$ is actually minimized for a much lower offset $1/Q_0=-7\times 10^{-9}$, which is consistent with the error bar. In other words, under the assumptions that the data follow a linear relation, the likelihood of the observed data given an arbitrary $1/Q_0$ is essentially maximized by the choice $1/Q_0\simeq 0$ (no systematic error). 
This is a further indication that the accuracy of the estimation of $1/Q$ is well within the measurement error bar.

\section{Measurements with higher coupling}
High coupling measurements were performed in a separate cooldown. The relative position of the cantilever with respect to the SQUID was carefully changed under the microscope and the system reassembled without other modifications.

The measurements were performed in similar way to the main run, but with a lower number of temperature points. Because of higher coupling the effective bandwidth of the cantilever noise was larger, leading in turn to a smaller error bar on the fitting parameters. Fig.~S4 shows three representative spectra. The data are again fitted by Eq.~(2). The $\chi^2$ is well within the $2 \sigma$ interval of the theoretical distribution for all spectra.

Fig.~S5 shows the $B$ parameter extracted from the fits as a function of $T/Q$.
An orthogonal linear fit of the data leads to $B_0=\left( 4.3 \pm 0.4 \right)\times 10^{-19}$ $\Phi_0^2/$Hz and $B_1= \left( 0.872 \pm 0.007 \right)\times 10^{-19}$  $\Phi_0^2/\left( \mathrm{nK} \cdot \mathrm{Hz} \right)$. The reduced $\chi^2$ with $4$ degrees of freedom is $0.21$, which falls within the $2 \sigma$ interval of the theoretical distribution. 
The coupling factor and the residual force noise inferred from $B_0$ and $B_1$ are reported in the second row of Table I.
\begin{figure}
\includegraphics{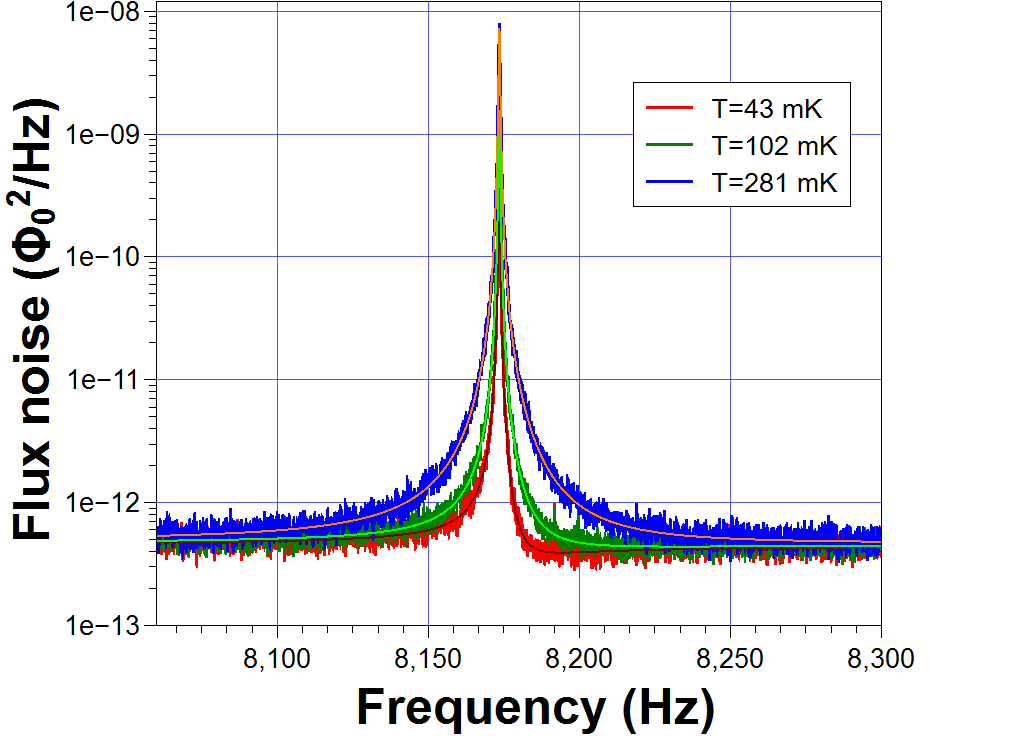}
\caption{Three representative spectra of the noise acquired in the high coupling run. The best fits with Eq.~(2) are also shown. \label{figS4}}  
\end{figure}
\begin{figure}
\includegraphics{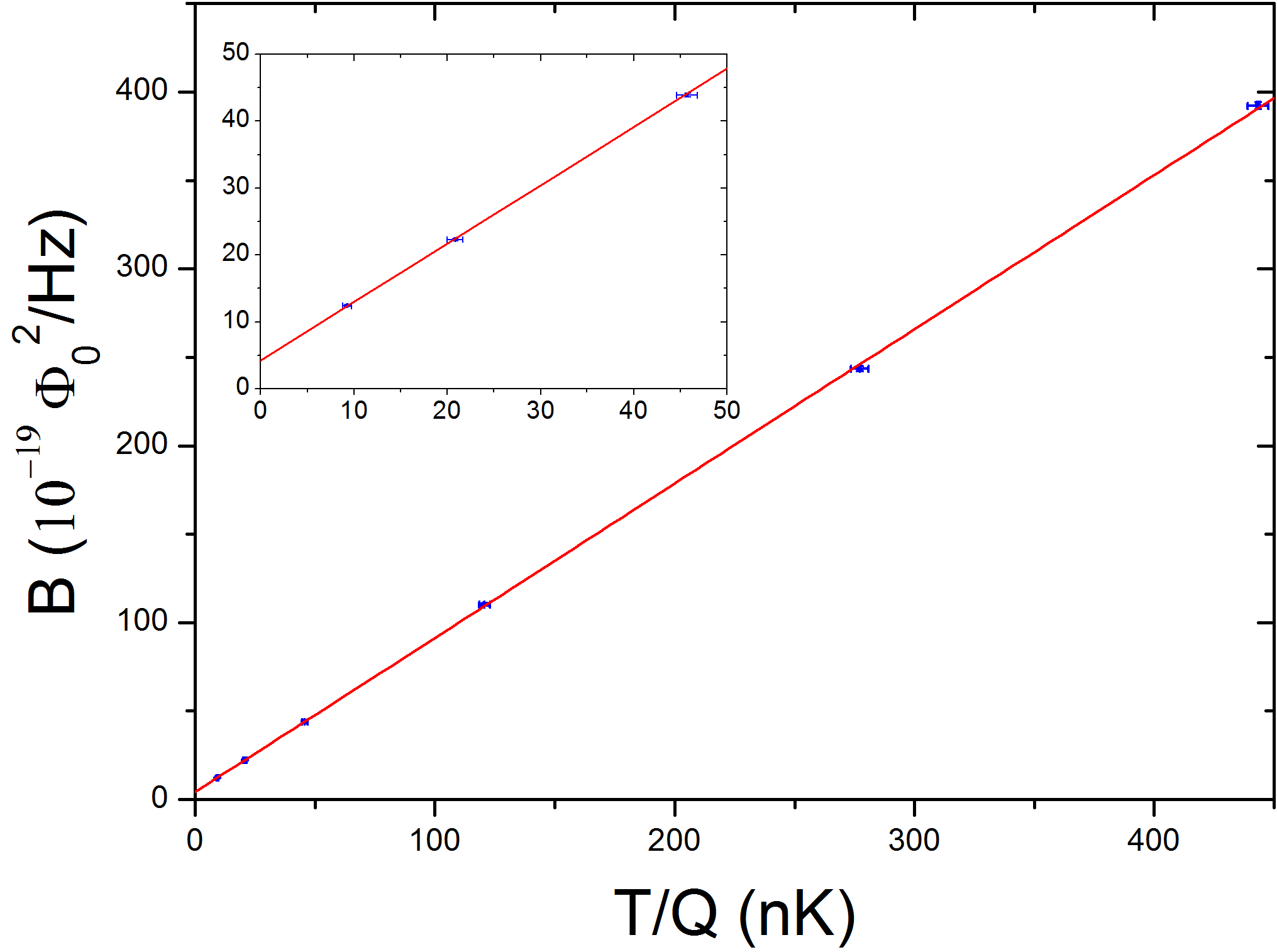}
\caption{$B$ parameter as a function of $T/Q$ for the high coupling dataset. The lowest points are zoomed in the inset. A linear fit is fully consistent with the data, yielding a finite intercept. \label{figS5}}  
\end{figure}

\section{Measurements with pulse tube on}
During the same cooldown of the main measurements, we have performed additional measurements without switching off the pulse tube cryocooler. Under normal operation, the high pressure pulses at frequency $\sim 1.5$ Hz generated by the pulse tube compressor are by far the strongest source of vibrational noise in our cryostat. 

Typically, we observe two different effects. On the one hand there is a direct generation of vibrational noise at the mixing chamber plate level, extending up to the $10$ kHz region, which can be directly measured by standard accelerometers. At the cantilever frequency the acceleration noise is less than $10^{-5}$ $g/\sqrt{\mathrm{Hz}}$ and our suspension system provides a factor $\sim  10^4$ attenuation. As the effective mass of our cantilever is $\sim  10^{-10}$ Kg, this translates into a force noise $\sim  1$ aN/$\sqrt{\mathrm{Hz}}$, thus comparable or lower than our residual measured force noise. The noise with pulse tube off is at least a factor of $10$ better (a factor $100$ in power).

On the other hand, very high vibrational noise levels are sometimes observed at the cantilever frequency due to nonlinear upconversion of low frequency noise. Upconversion is a poorly understood and rather unpredictable effect. It is highly nonstationary and threshold-like, with the noise at the resonator frequency which can vary by orders of magnitude, depending on the magnitude of the low frequency motion. We have evidence that upconversion is related either with soft thermal links or with the SQUID braided cable. We have been able to strongly reduce upconversion noise by implementing a passive magnetic damper in the suspension system, to reduce the low frequency motion, and by a proper clamping of the SQUID wiring. In particular, in the cooldown here considered, nonlinear upconversion was essentially absent even with the pulse tube on.

\begin{figure}
\includegraphics{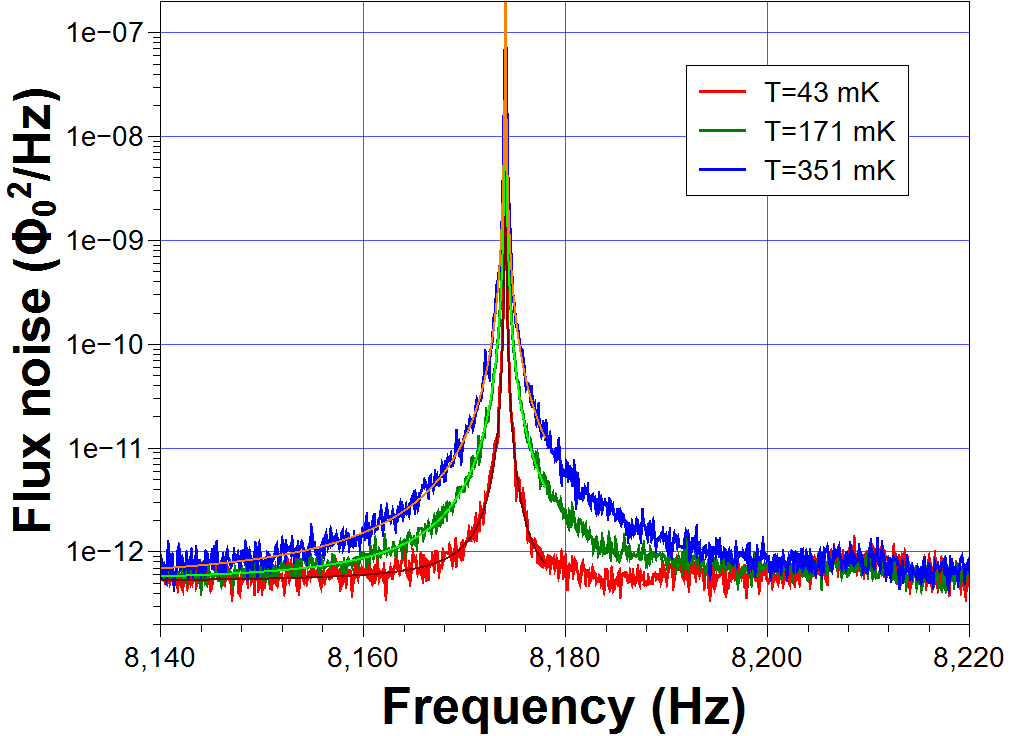}
\caption{Three representative spectra of the noise acquired with pulse tube on. Several bumps are apparent on the right tail. The best fit with Eq.~(2) are also shown. The fit is restricted to $f < 8178$ Hz and the parameters $A$ and $C$ are fixed to the values obtained with pulse tube off. \label{figS6}}  
\end{figure}
\begin{figure}
\includegraphics{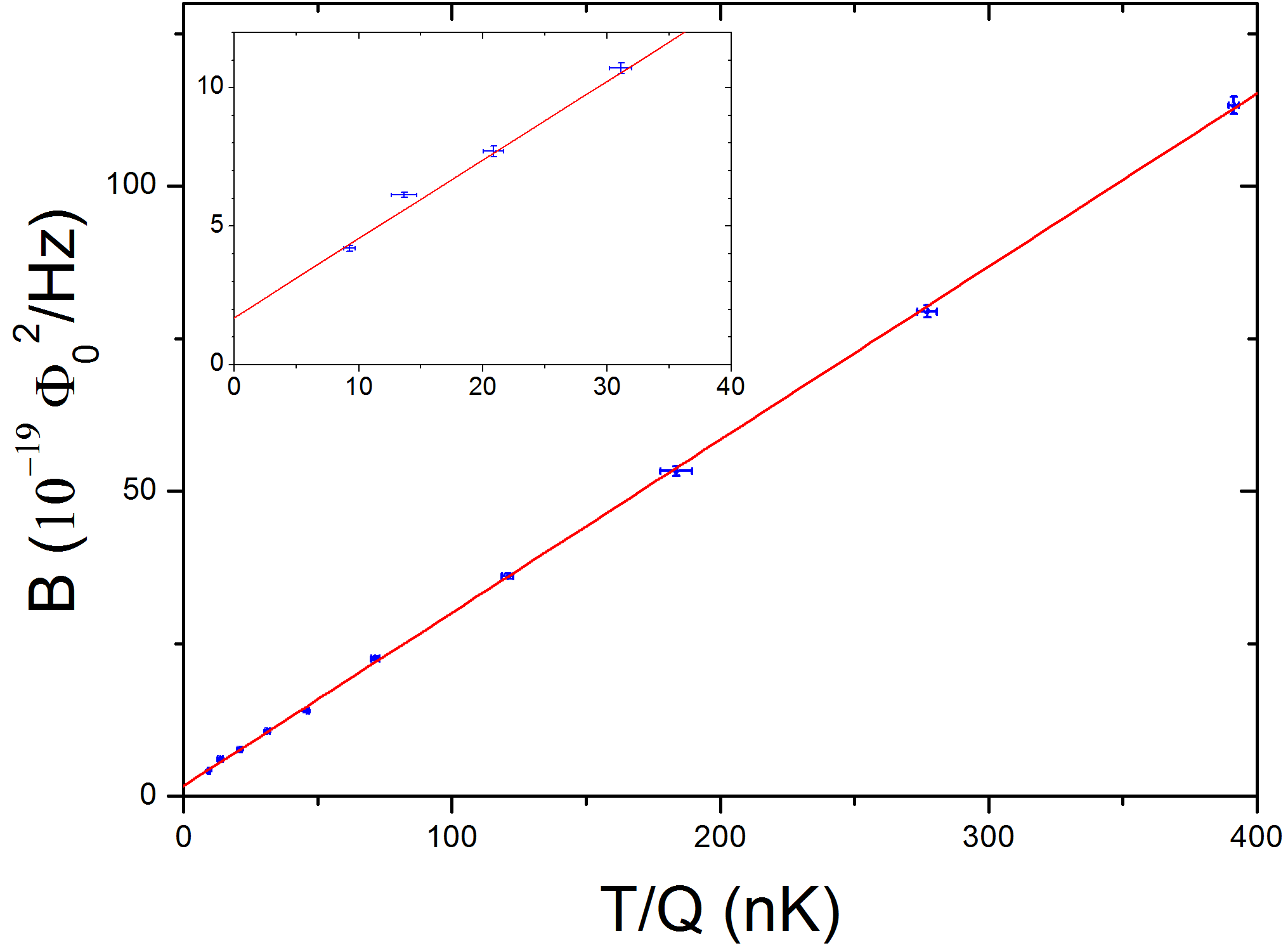}
\caption{$B$ parameter as a function of $T/Q$ for the measurements with pulse tube on. The lowest points are zoomed in the inset. A linear fit is again consistent with the data, yielding a finite intercept. \label{figS7}}  
\end{figure}
Fig.~\ref{figS6} shows three representative spectra acquired with the pulse tube on, with the same acquisition settings of the main measurements run. Several broad bumps are apparent on the right tail of the resonator peak, while the left tail is rather clean. A global fit yields a $\chi^2$ unacceptably high, which confirms the presence of coloured vibrational noise. However, we obtain acceptable $\chi^2$ by excluding a wide portion of the right tail from the fit, as shown in Fig.~\ref{figS6}. By applying the standard data analysis we obtain again a good linear behaviour of $B$ as a function of $T/Q$, as shown in Fig.~\ref{figS7}. The slope of the linear fit $B_1=(0.286 \pm 0.003) \times 10^{-19}$ $\Phi_0^2/\mathrm{Hz\cdot nK} $ is consistent with the one at pulse tube off, while the intercept $B_0=(1.71 \pm 0.13)\times 10^{-19}$ $\Phi_0^2/\mathrm{Hz} $ leads to a larger residual force noise (see Table I).

\section{Measurements with pump off}
Under pulse tube off operation, the stronger source of vibrational and acoustic noise is the roots mechanical pump which is employed to circulate the $^3$He-$^4$He mixture in the dilution refrigerator. We have tried to investigate whether the pump noise can be related to the observed cantilever excess noise. The measurement was performed during a separate cooldown with the high coupling setting at the lowest temperature $T=43$ mK.

Unfortunately, it is not possible to maintain a stable temperature for long time after switching off the circulation pump. The cooling power drops to zero very quickly and the temperature starts drifting after a time of the order of 1 minute. In contrast, we can easily operate the dilution refrigerator with pulse tube off up to half an hour while keeping the temperature of the mixing chamber actively stabilized.
\begin{figure}
\includegraphics{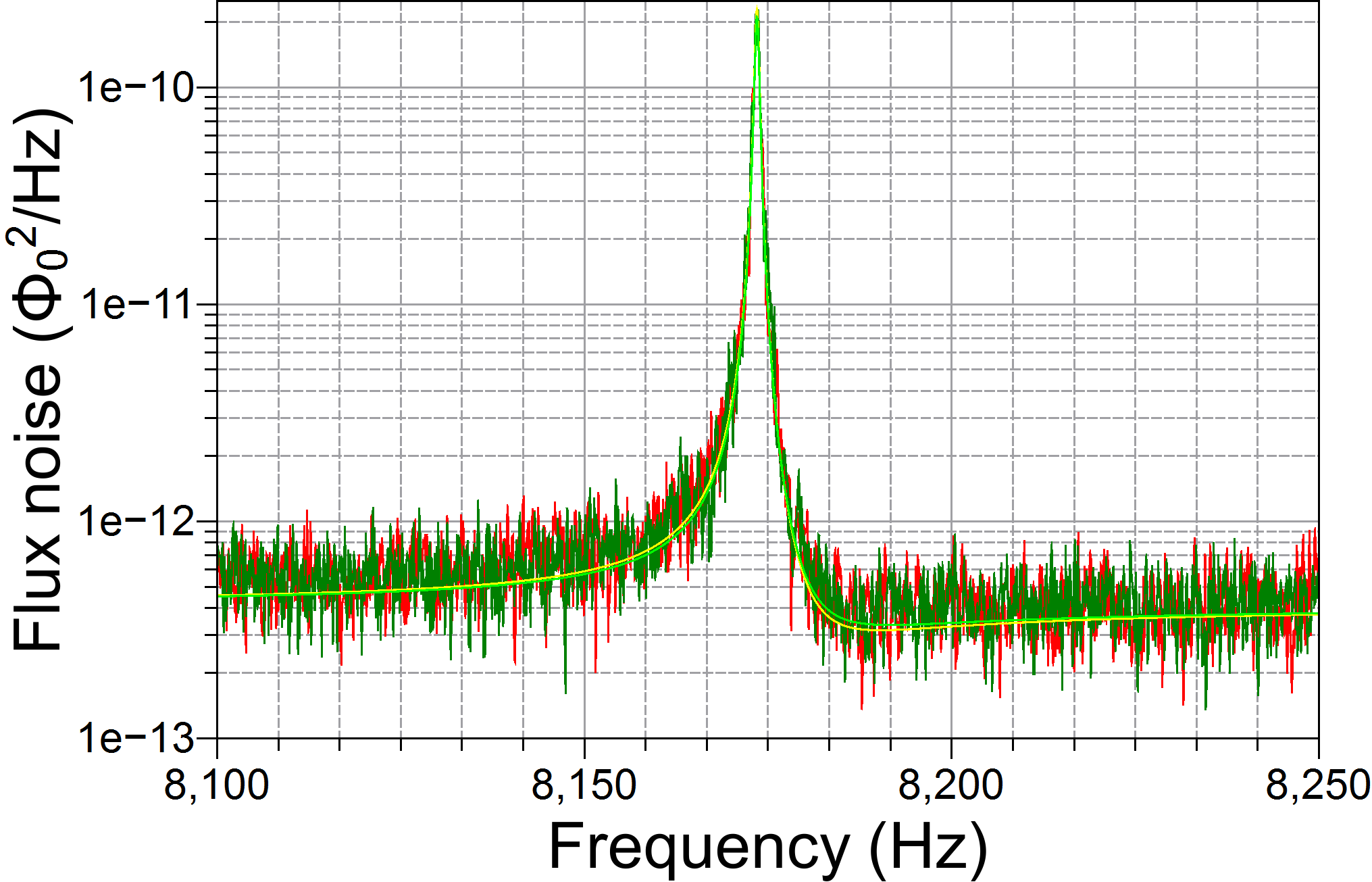}
\caption{Spectra acquired with circulation pump on (red line) and off (dark green line). The pulse tube is off. Both measurements are performed under the same conditions, with the same number of averages. The best fit with Eq.~(2) is also shown. No significant difference in the fitting parameters is observed. \label{figS8}}  
\end{figure}
In order to collect significant statistics while keeping a stable temperature, we switch off the circulation pump
for short periods (about 40 seconds), barely sufficient to wait for the low frequency suspension modes to relax and acquire one single data frame. Subsequently we switch the pump on, wait several minutes for the circulation to stabilize and then repeat the procedure. We collected a total of $12$ acquisitions. The averaged spectrum is then compared with a spectrum with circulation pump on. For a fair comparison, the spectrum with pump on is acquired with the same setting and the same number of averages.

The two spectra are shown in Fig.~S8, and can be hardly distinguished. The best fitting curves are also shown and are essentially coincident. The $B$ parameters resulting from the fits are $B=\left( 1.26 \pm 0.04 \right)\times 10^{-18}$ $\Phi_0^2/$Hz and $B=\left( 1.29 \pm 0.04 \right)\times 10^{-18}$ $\Phi_0^2/$Hz for the pump on and pump off case respectively. Therefore, there is apparently no significant effect of the circulation pump on the excess noise, which at this temperature contributes to about $30 \% $ of $B$.
\\
\section{Magnetic effects}
As the ferromagnetic microsphere is magnetized, an external magnetic field noise could be also held responsible for anomalous force driving the cantilever. Let us assume an environmental magnetic noise $B_n$ with direction along the cantilever length and negligible spatial dependence over the magnetic sphere volume. $B_n$ would generate a torque $\mu B_n$, where $\mu \simeq 5\times 10^{-9}$ J/T is the microsphere magnetic moment, which translates into an effective force noise $\mu B_n / l$, where $l$ is the effective length of the cantilever. Under these assumptions, the observed excess force noise would result from a magnetic field noise $B_n$ with spectral density $1\times 10^{-13}$ T/$\sqrt{\mathrm{Hz}}$. Such noise is typical of an unshielded environment at kHz frequency, but is unrealistically large for a shielded environment. The walls of the copper box hosting the cantilevers are about 20 times thicker than the penetration depth at the cantilever frequency, thus providing an attenuation of external magnetic fields by many orders of magnitude. Thermal magnetic noise from eddy currents in the walls or other elements inside the box is estimated to be largely negligible.

A related but distinct mechanism is given by fluctuations of the microsphere magnetization. The magnetic microsphere is at finite temperature, so there will be magnetization fluctuations, which will couple to the static magnetic field yielding a finite torque and force noise.  Magnetization fluctuations for a fully magnetized hard ferromagnet are expected to be very small, due to the very high anisotropy field. Experiments with rare-earth micromagnets have actually shown that at kHz frequency a larger effect is due to conductive eddy currents [36]. Along with the approach of [36] we estimate both effects to be many orders smaller than what is needed to explain the observed force noise. For instance, the eddy current dissipation in the microsphere can be calculated analytically and is 6 orders of magnitude smaller than the cantilever mechanical dissipation. 

However, we can also provide an experimental argument to rule out this mechanism. Magnetization fluctuations would behave as thermal force noise and the same mechanism would also appear as mechanical dissipation. In particular, both noise and dissipation would scale with the square of the external magnetic field. In a separate test we have have increased the external magnetic field by a factor 4 with respect to the earth field. We did not see any significant change of the quality factor, confirming that thermal magnetization fluctuations are likely not significant in our experiment.

\end{document}